\documentclass{article}

\usepackage{PRIMEarxiv}

\usepackage[utf8]{inputenc} 
\usepackage[T1]{fontenc}    
\usepackage{hyperref}       
\usepackage{url}            
\usepackage{booktabs}       
\usepackage{amsfonts}       
\usepackage{nicefrac}       
\usepackage{microtype}      
\usepackage{lipsum}
\usepackage{fancyhdr}       
\usepackage{graphicx}       
\graphicspath{{media/}}     

\usepackage{amsmath}

\title{An attention economy model of co-evolution between content quality and audience selectivity
}

\author{
  Masaki Chujyo\\
  Department of Systems Innovation, School of Engineering \\
  The University of Tokyo \\
  Japan\\
  \texttt{mchujyo@g.ecc.u-tokyo.ac.jp} \\
   \And
  Isamu Okada \\
  Department of Business Administration \\
  Soka University \\
  Japan\\
     \And
  Hitoshi Yamamoto \\
  Department of Business Administration \\
  Rissho University \\
  Japan\\
     \And
  Dongwoo Lim \\
  Media Studies Major, Faculty of Sociology \\
  Kansai University \\
  Japan\\
     \And
  Fujio Toriumi\\
  Department of Systems Innovation, School of Engineering \\
  The University of Tokyo \\
  Japan\\
}

\begin{document}
\maketitle

\begin{abstract}
Human attention has become a scarce and strategically contested resource in digital environments. Content providers increasingly engage in excessive competition for visibility, often prioritizing attention-grabbing tactics over substantive quality. Despite extensive empirical evidence, however, there is a lack of theoretical models that explain the fundamental dynamics of the attention economy. Here, we develop a minimal mathematical framework to explain how content quality and audience attention coevolve under limited attention capacity. Using an evolutionary game approach, we model strategic feedback between providers, who decide how much effort to invest in production, and consumers, who choose whether to search selectively for high-quality content or to engage passively. Analytical and numerical results reveal three characteristic regimes of content dynamics: collapse, boundary, and coexistence. The transitions between these regimes depend on how effectively audiences can distinguish content quality. When audience discriminability is weak, both selective attention and high-quality production vanish, leading to informational collapse. When discriminability is sufficient and incentives are well aligned, high- and low-quality content dynamically coexist through feedback between audience selectivity and providers’ effort. These findings identify two key conditions for sustaining a healthy information ecosystem: adequate discriminability among audiences and sufficient incentives for high-effort creation. The model provides a theoretical foundation for understanding how institutional and platform designs can prevent the degradation of content quality in the attention economy.
\end{abstract}

\keywords{Attention economy \and Adaptive feedback \and Replicator dynamics}

\section*{Introduction}
In today’s digital society, the attention economy has become a central paradigm for understanding online information ecosystems. Human attention functions as a limited resource subject to intense competition~\cite{davenport2001attention,falkinger2008limited,webster2016marketplace,bhargava2021ethics,pedersen2021political,vettehen2023attention,heitmayer2025second}.
This competition unfolds across digital media spaces, including news and social media platforms, video-sharing sites, and live-streaming environments~\cite{huberman2009crowdsourcing,weng2012competition,ciampaglia2015production,welbers2016news,leveille2018drawing,nixon2020business,watanabe2021attention,smith2021pay,mao2022effectiveness,liang2022end,dai2023effect,wollborn2023competitive}, where user-generated and professional content continually vie for visibility.
As digital platforms lower the barriers to content creation, the volume of online content has surged. This expansion has intensified competition for user attention and enabled even low-quality material to spread widely~\cite{lasser2022social,mosleh2025divergent}.
This imbalance between abundant content and users’ limited capacity to process it highlights the fundamental scarcity of attention in online environments.

A widely observed pattern in such competitive environments is the proliferation of low-effort content, including clickbait, optimized to exploit platform algorithms and maximize short-term engagement~\cite{biyani20168,chen2015misleading,lischka2023clickbait}.
Economic models show that competition for fragmented attention can depress information quality~\cite{chen2023competition} and may even incentivize misinformation when short-term gains outweigh long-term credibility~\cite{amini2025media}.
Recent advances in generative AI have amplified these pressures. They enable the rapid production of large volumes of text, images, and videos at minimal cost~\cite{wang2023survey,sandrini2023generative,knott2024ai}.
Although these strategies may be economically rational for individual providers, they collectively threaten the long-term viability of high-quality content and the informational health of online ecosystems.
These observations raise a fundamental question: under what conditions can high-quality content remain sustainable in environments where limited attention is strategically distributed?

To address this question, it is essential to recognize that online information environments are shaped by the strategic adaptations of both content providers and consumers.
Providers adjust their production strategies in response to competitive pressures and anticipated visibility. Consumers allocate their attention based on perceived credibility and content quality.
These reciprocal adjustments create feedback loops that shape the evolving relationship between content production and attention allocation~\cite{huberman2008social,zhu2020content,krebs2025model}.
Previous studies have largely examined either collective attention patterns or individual attention decay~\cite{salganik2006experimental,wu2007novelty,
weng2012competition,ojer2025modelingindividualattentiondynamics}.
However, these approaches do not capture how providers and consumers jointly adapt under attention constraints.
A theoretical framework that integrates their coadaptive behaviors and clarifies how such constraints shape long-term evolutionary outcomes is still lacking.
In particular, many existing models treat attention allocation and content quality as exogenous, which limits their ability to capture the feedback between the two. 
As a result, how limited attention shapes the coevolution of quality provision and consumer behavior remains insufficiently understood.

Building on these observations, we develop a theoretical framework that explains when high-quality content can persist under limited attention.
Unlike previous approaches that examine attention flows or individual behavior in isolation, our model jointly captures the strategic adaptation of content providers and consumers under explicit attention constraints.
We construct a two-population evolutionary game with behavioral microfoundations, in which interactions under finite attention capacity generate a piecewise-smooth replicator system derived through a mean-field approximation.
This formulation enables us to analytically identify the parameter thresholds that separate collapse, boundary, and coexistence regimes, providing a unified account of how attention constraints shape the long-term evolutionary dynamics of content quality.

Our analysis reveals that limited attention induces a fundamental divide in content dynamics: without sufficient discriminability, systems generically collapse into low-quality convergence, whereas boundary outcomes and stable coexistence arise only under specific combinations of attention capacity and providers’ incentive strength. These evolutionary patterns clarify when high-effort content can persist in attention-driven environments and offer theoretical guidance for designing platforms that maintain informational quality. We formalize these mechanisms by introducing the evolutionary framework on which our analysis is based.

\section*{Results}

\subsection*{Attention Economy Model}

\begin{figure}[t]
    \centering
    \includegraphics[width=0.8\linewidth]{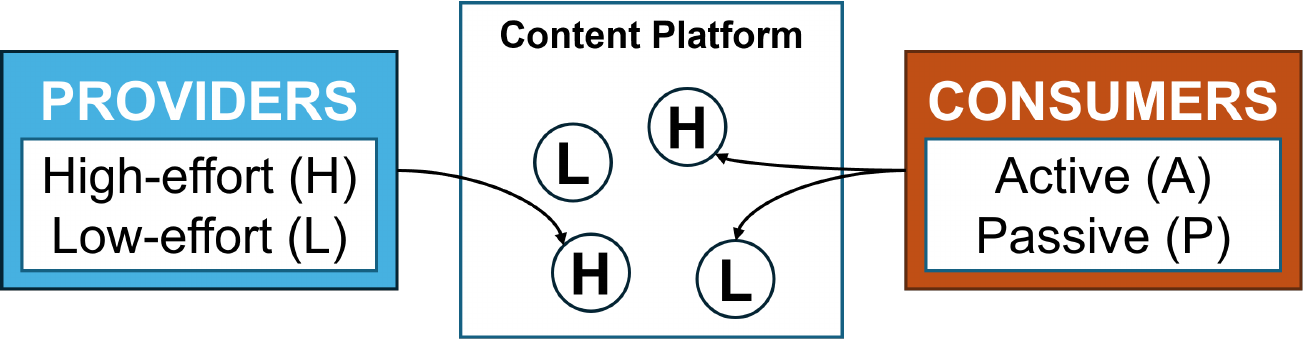}
    \caption{Conceptual diagram of the attention economy model. Providers invest high (H) or low (L) effort in content creation, while consumers allocate limited attention $(\sigma)$ through active (A) or passive (P) viewing. Their strategic feedback determines the evolution of content quality and attention distribution.}
    \label{fig:illustrate}
\end{figure}

To formalize the coadaptive mechanisms discussed above, we construct a minimal evolutionary framework that captures how content quality and audience attention coevolve under limited cognitive resources. Digital environments are characterized by intense competition for visibility, which makes the incentives of providers and consumers inherently interdependent.
The structure of the model is summarized in Fig.~\ref{fig:illustrate}.

Content providers choose between two strategies. Under the high-effort strategy $H$, they produce high-quality content at a cost $c_H>0$. Under the low-effort strategy $L$, they generate lower-quality content with no additional cost. We consider a fixed population of $m$ providers, each producing one item of content per period, regardless of strategy. Providers receive a reward $r>0$ per view, so the payoff advantage of high-effort production depends on how consumers allocate attention across available items. The reward $r$ is identical for high- and low-quality items, reflecting platform environments, such as video-sharing platforms, where provider revenue is largely determined by view counts rather than intrinsic content quality.

Consumers also choose between two strategies. Under the active strategy $A$, they search for and selectively consume high-quality items while paying a per-period cost $c_A>0$ that represents search and evaluation effort. Under the passive strategy $P$, they allocate their attention budget uniformly across all items produced. We consider a fixed population of $n$ consumers, each with a limited attention capacity $\sigma \in (0,1]$, which allows them to consume only a fraction $\sigma$ of items produced in each period. Consumer benefits from each item are additively separable.

Attention is treated as a scarce cognitive resource. The parameter $\sigma$ formalizes this constraint by limiting each consumer’s attention budget, forcing providers to compete for inclusion within a narrow attention window. As a result, visibility becomes a key determinant of provider incentives and directly reflects consumer selection behavior. A key implication of this assumption is that attention scarcity enters payoffs directly, creating a qualitative shift in incentives depending on whether high-quality supply is sufficient to saturate consumers’ attention budgets. The boundary $x=\sigma$ thus marks the point at which active consumers can fully satisfy their attention budgets using only high-quality items, eliminating any residual allocation to low-quality content and altering the payoff structure accordingly.

\begin{table}[t]
\centering
\caption{Summary of model parameters.}
\label{tab:parameters}
\begin{tabular}{ll}
\toprule
\textbf{Symbol} & \textbf{Description} \\
\midrule
$m$ & Number of content providers \\
$n$ & Number of content consumers \\
$\sigma$ & Attention capacity (fraction of attendable items) \\
$b_H$ & Benefit obtained from a high-quality item \\
$b_L$ & Benefit obtained from a low-quality item \\
$c_H$ & Cost of adopting the high-effort strategy\\
$c_A$ & Cost of adopting the active viewing strategy \\
$r$ & Reward per view received by providers \\
$x(t)$ & Fraction of high-effort providers at time $t$ \\
$y(t)$ & Fraction of active consumers at time $t$ \\
\bottomrule
\end{tabular}
\end{table}

Given these incentives, the composition of strategies in each population evolves over time. 
Strategies adjust in proportion to their payoff differences, with higher-payoff strategies becoming more common. 
This replicator-like adjustment creates feedback between content quality and consumer selectivity.

Interactions in each period proceed as follows:
\begin{enumerate}
    \item Providers produce one item according to their chosen strategy.
    \item Consumers select $m\sigma$ items. Active consumers allocate their attention entirely to high-quality items when these are available in sufficient quantity. When the supply of high-quality items is insufficient, they consume all available high-quality items and then randomize over low-quality items to fill their remaining attention slots. Passive consumers sample uniformly from all items.
    \item Providers receive reward $r$ per view.
    \item Consumers obtain benefit $b_H$ from high-quality and $b_L$ from low-quality items, with $b_H > b_L > 0$.
\end{enumerate}
The model parameters are listed in Table \ref{tab:parameters}. 

Let $x(t)$ denote the fraction of high-effort providers and $y(t)$ the fraction of active consumers at time $t$. These state variables summarize the population-level state. We capture the resulting feedback through a two-population replicator framework, with the full mathematical formulation provided in the Materials and Methods section.

To clarify how attention constraints shape the coevolution of content quality and viewing behavior, we analyze the qualitative behavior of this replicator system. We first characterize all equilibria and derive the parameter thresholds that determine their stability. 
We then perform numerical integration of the system to examine trajectories and basins of attraction, illustrating how the system responds to changes in attention capacity and incentive strength.

Across these analyses, the dynamics organize into three characteristic regimes that govern long-run outcomes of attention and quality. In the collapse regime, both selective attention and high-effort provision vanish, and the system converges to a low-quality, non-selective state. In the boundary regime, consumers remain selective but providers exert only intermediate effort, leading to partially improved yet fragile information environments. In the coexistence regime, high-effort providers and selective consumers persist together with damped oscillations, maintaining quality through mutual adaptation. In the following sections, we characterize the parameter conditions under which each regime arises and analyze their dynamical and policy implications.

\subsection*{Analytical characterization of stable equilibria}
\label{subsec:equilibrium_stability}

We begin by analyzing the stability of the three equilibria identified in the replicator system:    
(i) the collapse equilibrium at the origin,  
(ii) the boundary equilibrium on \(y=1\) with \(0 < x^* < 1\), and  
(iii) an interior coexistence equilibrium in which both high-effort providers and active consumers persist.
As summarized in Table~\ref{tab:equilibria_summary}, the stability of these equilibria is governed by the interplay between attention capacity, the cost of active consumption, and the effective reward for high-effort provision.

\begin{table*}[t]
\centering
\caption{Summary of equilibrium types and their stability conditions in the attention-quality dynamics.}
\label{tab:equilibria_summary}
\begin{tabular}{lccc}
\toprule
Type of equilibrium & Fixed point $(x^*,y^*)$ & Stability condition \\ 
\midrule
Collapse equilibrium &
$(0,0)$ &
Always stable \\[3pt]

Boundary equilibrium &
$\left(\dfrac{r n \sigma}{c_H},\,1\right)$ &
$\begin{cases}
r n \sigma< c_H < r n,\\
\dfrac{c_A}{m\sigma(b_H-b_L)}+\dfrac{r n\sigma}{c_H} < 1
\end{cases}$ \\[8pt]

Coexistence equilibrium &
$\left(1-\dfrac{c_A}{m\sigma(b_H-b_L)},\,\dfrac{c_H}{r n\sigma}\!\left(1-\dfrac{c_A}{m\sigma(b_H-b_L)}\right)\right)$ &
$\begin{cases}
m\sigma(1-\sigma)(b_H-b_L) > c_A,\\[3pt]
\dfrac{c_A}{m\sigma(b_H-b_L)}+\dfrac{r n\sigma}{c_H} > 1
\end{cases}$ \\[8pt]
\bottomrule
\end{tabular}
\end{table*}

The origin \((x, y)=(0,0)\) is always locally stable, representing a collapse regime where both selective attention and high-effort provision disappear.  
In particular, this equilibrium becomes the only attractor when the following condition holds:
\begin{equation}
    m\sigma(1-\sigma)(b_H-b_L) < c_A .
\end{equation}
The derivation of this threshold is provided in the SI Appendix.
Intuitively, the term $\sigma(1-\sigma)$ indicates that discrimination is weakest when attention is either too scarce or too dispersed, making selective viewing unprofitable and preventing high-quality content from gaining traction.
In this parameter region, no nontrivial equilibria exist, and all trajectories converge to the origin regardless of initial conditions.
In such environments, consumers find it unprofitable to engage in selective attention, and low-effort content dominates in the steady state.

In contrast, when
\(m\sigma(1-\sigma)(b_H-b_L) > c_A\),
nontrivial equilibria become possible. 
Which equilibrium stabilizes depends on the ratio of the effective reward to the production cost, \(r n \sigma / c_H\). 
If
\begin{equation}
    \frac{c_A}{m\sigma(b_H-b_L)}+\frac{r n\sigma}{c_H} > 1,
\end{equation}
the interior coexistence equilibrium is locally stable. 
In this regime, both high-effort providers and active consumers persist at positive levels, sustaining quality endogenously through mutual reinforcement. 
In contrast, if the opposite inequality holds,
\begin{equation}
    \frac{c_A}{m\sigma(b_H-b_L)}+\frac{r n\sigma}{c_H} < 1,
\end{equation}
the boundary equilibrium becomes stable: all consumers remain active (\(y\to1\)), but providers exert only intermediate effort. 
This state represents a partially functional but not fully self-sustaining information environment.

\begin{figure}[thp]
    \centering
    \includegraphics[width=0.9\linewidth]{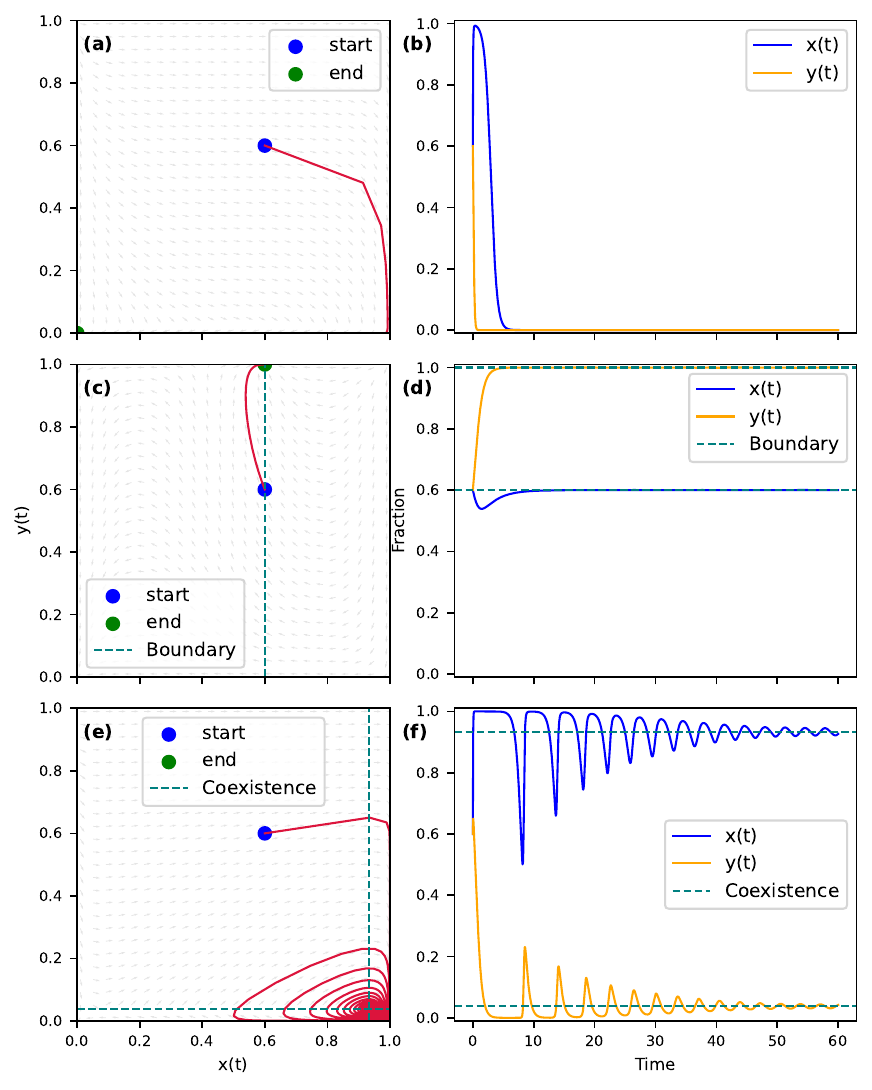}
    \caption{Representative trajectories of the replicator dynamics under three parameter regimes. Parameters were chosen to satisfy the corresponding stability conditions: (a,b) Collapse: $m=20$, $n=100$, $\sigma=0.5$, $r=1$, $b_H-b_L=1$, $c_A=10$, $c_H=2$.  (c,d) Boundary: $m=1$, $n=10$, $\sigma=0.5$, $r=0.12$, $b_H-b_L=10$, $c_A=1$, $c_H=1$.  (e,f) Coexistence: $m=20$, $n=100$, $\sigma=0.5$, $r=1$, $b_H-b_L=3$, $c_A=2$, $c_H=2$. The green dashed lines mark the  equilibrium.}
    \label{Fig2}
\end{figure}

These analytical results indicate two hierarchical thresholds that shape the sustainability of quality under limited attention. The first threshold, \(m\sigma(1-\sigma)(b_H-b_L)=c_A\), separates collapsed from non-collapsed systems and depends solely on consumers’ ability to distinguish between high- and low-quality content. The second threshold, determined by the balance between reward and cost parameters, divides the non-collapsed region into boundary and coexistence regimes. Together, they show that maintaining high-quality information requires not only sufficient discriminability among consumers but also an adequate incentive structure for providers.

\subsection*{Dynamic behavior across three regimes}
\label{subsec:regime_dynamics}
Next, we examine how system trajectories approach these equilibria under varying parameter regimes.
Figure~\ref{Fig2} presents representative vector fields and time evolutions for the three regimes: collapse, boundary, and coexistence. 
These trajectories were generated by numerically integrating the replicator equations with a standard adaptive Runge–Kutta method (RK45), using parameter values chosen to satisfy the stability conditions of each regime.

When the active-consumption cost \(c_A\) is high or the quality gap \(b_H - b_L\) is small, consumers lack sufficient motivation to search for high-quality content. Selective viewing fails to emerge (\(y \to 0\)), and providers lose incentives to exert high effort (\(x \to 0\)), leading both populations to a state dominated by low-quality content and non-selective behavior.

In the boundary regime, consumers sustain selective attention (\(y \to 1\)), but the effort level of providers remains moderate, with \(x\) converging to an intermediate value. This regime represents an environment where demand for quality exists, yet the reward for high-effort provision is inadequate to sustain full-quality production.

In the coexistence regime, both providers and consumers persist at positive equilibrium levels (\(x > 0, y > 0\)), yet convergence toward this steady state is not monotonic. The trajectories exhibit mild oscillations around the equilibrium, reflecting feedback between supply and selection. When the share of low-quality content increases, consumers become more selective (\(y\) rises), which restores incentives for high-effort creation and raises \(x\). As high-quality content becomes abundant, the need for active searching decreases, and \(y\) gradually falls. This in turn reduces \(x\), generating a slow cyclical adjustment. Because these adjustments weaken as the system approaches the fixed point, the feedback loses strength over time, producing damped rather than persistent oscillations.

This damped cycle reflects adaptive consumer behavior: selectivity increases only when necessary, allowing consumers to minimize active-consumption costs while supporting the persistence of high-quality supply. Thus, the coexistence regime represents a dynamically maintained state in which high-effort provision and selective attention reinforce one another.

\subsection*{Basins of attraction across regimes}
\label{subsec:basins}
Having examined the typical trajectories within each regime, we next investigate how their basins of attraction vary across parameter settings. The size and structure of these basins differ substantially between the boundary and coexistence regimes.

\begin{figure}[th]
\centering
\includegraphics[width=0.8\linewidth]{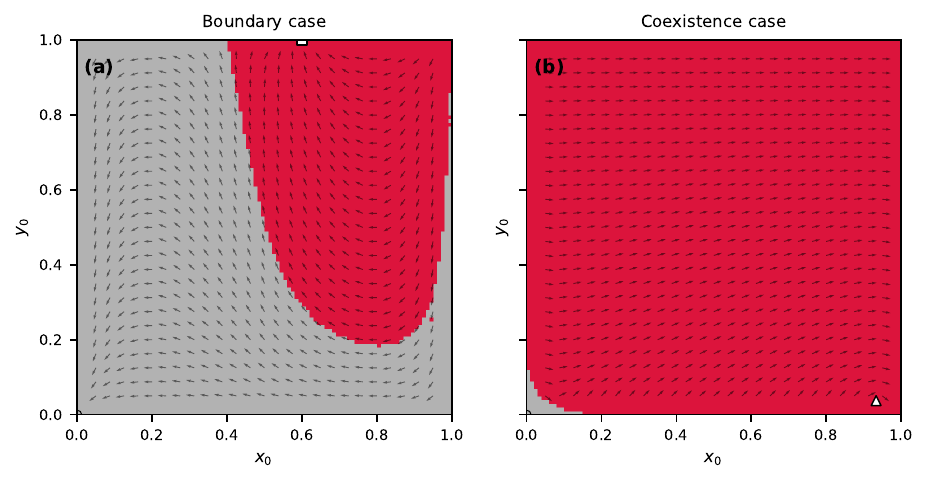}
\caption{Basins of attraction under two parameter regimes. Left: parameter setting where the boundary equilibrium is stable. Right: setting where the coexistence equilibrium is stable. Red regions indicate initial conditions converging to non-collapsed equilibria, whereas gray regions indicate collapse. Background arrows show the replicator vector field. Basins were computed on a 100 $\times$ 100 grid of initial conditions.}
\label{fig:basins}
\end{figure}

Figure~\ref{fig:basins} shows the final equilibrium reached from systematically varied initial conditions \((x_0, y_0)\). Red regions denote initial conditions that converge to the coexistence or boundary regime, while gray regions correspond to trajectories converging to the collapse regime. Arrows in the background indicate the vector field of the replicator dynamics.

When the boundary equilibrium is stable (left panel), the region of initial conditions leading to coexistence is limited. Without a sufficiently large initial fraction of high-effort providers or active consumers, the system collapses to \((x, y) \to (0, 0)\). This indicates that the boundary regime is highly sensitive to initial conditions: quality and selectivity must already be present at appreciable levels for the system to avoid degeneration.

In contrast, when the internal coexistence equilibrium is stable (right panel), the basin of attraction expands substantially.  
Even when both \(x\) and \(y\) start from low levels, the feedback between effort and attention drives the system back toward coexistence.  
This result indicates that the coexistence regime is robust across a wide range of initial conditions, whereas the boundary regime emerges only when selective attention or high-effort provision is already prevalent.

\subsection*{Determinants of collapse and policy implications}
\label{subsec:drivers_policies}

The key bifurcation of the system separates the collapse regime, where both strategies vanish, from the non-collapsed regimes, where at least partial coexistence of high- and low-quality content persists.
This transition primarily depends on consumers’ ability to discriminate between quality levels, which is characterized by $m\sigma(1 - \sigma)(b_H - b_L) = c_A$.
Here, \(\sigma(1 - \sigma)\) captures a nonlinear effect of attention capacity: discrimination weakens when attention is either too scarce or too dispersed.  
If this balance between discriminability and active-consumption cost fails to exceed this threshold, selective engagement does not emerge, and high effort cannot be sustained. In this case, the system collapses regardless of initial conditions.

\begin{figure}[th]
\centering
\includegraphics[width=0.8\linewidth]{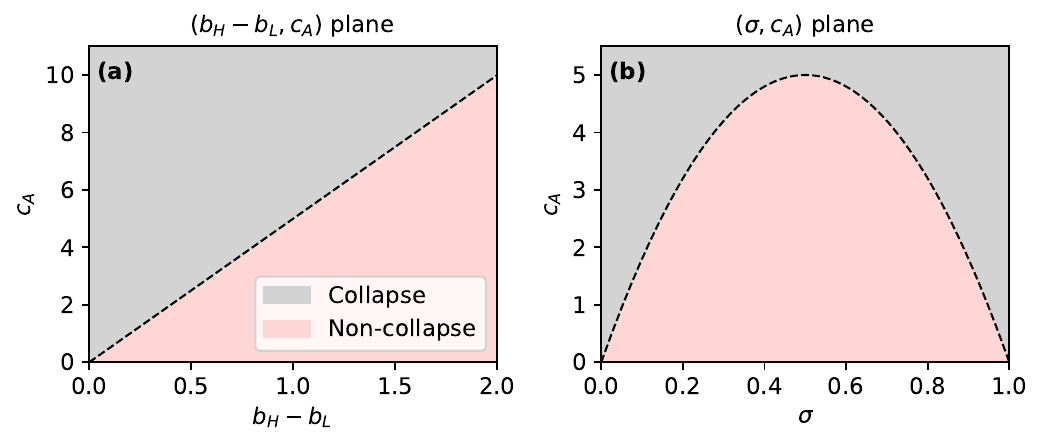}
\caption{Regimes of collapse and non-collapse. (a) Regimes on the \((b_H - b_L, c_A)\) plane for fixed \(\sigma = 0.5\). The dashed line \(c_A = m\sigma(b_H - b_L)(1 - \sigma)\) separates collapse (gray, above) from non-collapse (white, below). (b) Regimes on the \((\sigma, c_A)\) plane for fixed \(b_H - b_L = 1\). The non-collapsed region is widest around \(\sigma \approx 0.5\) and shrinks as \(\sigma \to 0\) or \(\sigma \to 1\).}
\label{fig:phase_regimes}
\end{figure}

Figure~\ref{fig:phase_regimes} shows this boundary in the \((b_H - b_L, c_A)\) and \((\sigma, c_A)\) planes.  
A larger quality gap or lower active-consumption cost reduces the collapse region.  
The non-collapsed region is maximized near \(\sigma = 0.5\) and shrinks as \(\sigma \to 0\) or \(\sigma \to 1\).  
In smaller markets with fewer providers (\(m\) small), preventing collapse requires stronger support for discrimination (reducing \(c_A\)) or a larger quality gap.

Importantly, this collapse boundary is independent of the reward parameters \((r, c_H, n)\) that govern providers’ incentives.  
The first-order condition for sustaining quality lies in the consumers’ motivation to differentiate content.  
Once this condition is met, the specific equilibrium that stabilizes within the non-collapsed domain is determined by the ratio \(r n \sigma / c_H\): higher ratios favor the coexistence equilibrium, whereas lower ratios yield boundary-type stability.

These results suggest a two-step design principle for sustaining high-quality digital ecosystems.  
First, collapse must be avoided by ensuring sufficient discriminability.  
This can be achieved by (i) reducing active-consumption cost \(c_A\) through improved search, labeling, or recommendation systems,  
(ii) enlarging the quality gap \(b_H - b_L\) by suppressing low-quality exposure or penalizing manipulative content, and  
(iii) optimizing attention capacity \(\sigma\) by limiting excessive information exposure.  
Second, once collapse is avoided, maintaining coexistence requires increasing the effective reward ratio \(r n \sigma / c_H\) by improving incentives for high-effort production or reducing creation costs.  

Overall, the model demonstrates that sustainable content quality
does not rely on favorable initial conditions but on an institutional balance between discriminability and incentive design.

\section*{Discussion}
\label{sec:discussion}

To understand how content quality and audience attention coevolve under limited attention, we developed a minimal theoretical framework that captures the strategic feedback between providers and consumers.  
The model reveals that content dynamics fall into three regimes separated by a fundamental bifurcation determined by \(m\sigma(1 - \sigma)(b_H - b_L) = c_A\).  
This threshold clarifies the role of consumers’ discriminability, with \(\sigma(1 - \sigma)\) indicating that both excessive and insufficient attention undermine quality selection.  
Thus, simply increasing exposure does not necessarily improve the content environment.

Our contribution is to show that these dynamics emerge from a simple coadaptive mechanism linking selective attention and content effort under explicit attention constraints.

Each regime has a natural interpretation.  
In the collapse regime, low-quality and non-selective behaviors dominate, resembling environments saturated with repetitive, manipulative, or clickbait content~\cite{biyani20168,chen2015misleading,lischka2023clickbait}.  
The coexistence regime represents a dynamic balance in which temporary declines in quality heighten selectivity, restoring providers’ incentives and generating self-correcting oscillations.  
The boundary regime arises when consumers remain selective but high-effort production is only moderately profitable, a situation characteristic of expert or niche domains.
Our basin analysis shows that coexistence is robust to a wide range of initial conditions, whereas boundary outcomes require already favorable levels of selectivity or quality. This asymmetry highlights the importance of interventions that support discriminability before content quality deteriorates.

A related implication concerns recent developments in generative AI. By enabling the rapid creation of large amounts of content, such tools may increase the volume of material competing for users’ attention. This can heighten the cognitive effort required to identify reliably high-quality items, effectively raising the functional cost $c_A$ in our model. While this remains a tentative interpretation, it highlights a potential mechanism through which content proliferation may move platforms closer to the collapse boundary when discrimination does not improve accordingly.

These findings offer a two-stage framework for policy and platform design. First, avoiding collapse requires sustaining sufficient discriminability, which can be achieved by lowering cognitive search costs, enlarging the quality gap, or optimizing attention capacity. Importantly, the system performs best in an intermediate range where \(\sigma(1 - \sigma)\) is maximized; increasing attention indiscriminately is not always beneficial.  
Second, once collapse is avoided, stabilizing coexistence requires improving the effective reward ratio \(r n \sigma / c_H\) through transparent incentives, lower production costs, or appropriately regulated content throughput. Sustaining content quality thus depends less on initial conditions than on an institutional balance between discriminability and incentives.

The present model abstracts many features of real digital ecosystems. It assumes homogeneous consumers and providers, fixed market size, and simplified functional forms for discrimination and rewards. Relaxing these assumptions could reveal richer dynamics such as multistability or path dependence. Empirically evaluating the model’s predictions will require data on selective engagement and content production across platforms. Future extensions may also incorporate cases where low-quality content yields subjective or identity-based utility, such as misinformation, sensationalism, or partisan narratives. Integrating media literacy, trust, and informational health into the model could further clarify how attention and quality interact under algorithmic amplification~\cite{guess2020digital,kozyreva2024toolbox,toriumi2024informational}.

Overall, the model indicates that sustaining high-quality content hinges not on favorable initial conditions but on preserving discriminability while maintaining incentives for high-effort creation.

\section*{Material and Methods}
\subsection*{Replicator formulation}
We now develop the analytical formulation of the evolutionary dynamics.  
Let \(x(t)\) denote the fraction of high-effort providers and \(y(t)\) the fraction of active consumers.  
The incentive to invest in effort increases with the prevalence of active consumers, while the value of being active rises with the share of high-quality providers.  
This bidirectional feedback generates a coevolutionary dynamic process in the pair \((x(t), y(t))\).  

In each period, every provider produces one item, and each consumer can attend to a fraction \(\sigma\) of the items produced.  
We analyze the system at the population level by adopting a mean-field framework, in which providers and consumers are randomly matched and individual fluctuations are smoothed by averaging.
The parameter \(\sigma\) thus determines whether the system operates in an attention-saturated or attention-limited regime, depending on whether the fraction of high-quality content $x$ is large enough to meet the attention capacity $\sigma$.
Accordingly, the payoff structure differs between these regimes, leading to a piecewise-defined form of the expected payoffs that captures the transition between attention-saturated and attention-limited environments.
This distinction arises because active consumers may or may not be able to fill their attention budgets with high-quality items, causing the payoff differences to take distinct functional forms across regimes.

Let $\pi$ denote the expected payoff for each strategy under the mean-field approximation.  
Subscripts \(H\) and \(L\) refer to high- and low-effort providers, respectively, while \(A\) and \(P\) denote active and passive consumers, respectively. Superscripts "prov" and "cons" denote the provider and consumer populations.
\paragraph{(i) Case 1: high-quality supply sufficient (\(x \ge \sigma\)).}
\begin{equation}
\begin{aligned}
\pi_H^{\mathrm{prov}}(x,y) &= -c_H + r n \sigma\left((1-y) + \frac{y}{x}\right),\\
\pi_L^{\mathrm{prov}}(x,y) &= r n \sigma(1-y),\\
\pi_A^{\mathrm{cons}}(x,y) &= -c_A + m\sigma b_H,\\
\pi_P^{\mathrm{cons}}(x,y) &= \sigma m\left(x b_H + (1-x)b_L\right).
\end{aligned}
\end{equation}
\paragraph{(ii) Case 2: high-quality supply insufficient (\(x < \sigma\)).}
\begin{equation}
\begin{aligned}
\pi_H^{\mathrm{prov}}(x,y) &= -c_H + r n\left(\sigma(1-y) + y\right),\\
\pi_L^{\mathrm{prov}}(x,y) &= r n\left[\sigma(1-y) + y\frac{\sigma-x}{1-x}\right],\\
\pi_A^{\mathrm{cons}}(x,y) &= -c_A + m\left(x b_H + (\sigma-x)b_L\right),\\
\pi_P^{\mathrm{cons}}(x,y) &= \sigma m\left(x b_H + (1-x)b_L\right).
\end{aligned}
\end{equation}
These expressions are derived by conditioning on whether active consumers can fully satisfy their attention budgets with high-quality items.  
Detailed derivations are provided in the SI Appendix.

Both providers and consumers adapt their strategies through a finite rationality imitation process. Individuals observe the payoffs of others in their population and switch strategies with a probability proportional to the payoff difference. 
Thus, the evolution of the population shares is described by a two-population replicator system:
\begin{equation}
\begin{aligned}
\dot{x} &= x(1-x)\left[\pi_H^{\mathrm{prov}}(x,y) - \pi_L^{\mathrm{prov}}(x,y)\right],\\
\dot{y} &= y(1-y)\left[\pi_A^{\mathrm{cons}}(x,y) - \pi_P^{\mathrm{cons}}(x,y)\right].
\end{aligned}
\end{equation}
As the payoff functions are piecewise defined at \(x = \sigma\), the resulting vector field constitutes a piecewise-smooth dynamical system.  
Substituting the regime-specific payoff functions into the replicator equations yields the following piecewise-smooth system:
\begin{align}
\dot{x} &=
\begin{cases}
(1 - x)\left( r n \sigma y - c_H x \right), & (\sigma \leq x), \\
x \left( r n y(1 - \sigma) + c_H(x - 1) \right), & (\sigma > x),
\end{cases} 
\label{eq:dxdt_piecewise}
\\
\dot{y} &=
\begin{cases}
y(1 - y)\left( b m \sigma (1 - x) - c_A \right), & (\sigma \leq x), \\
y(1 - y)\left( b m x (1 - \sigma) - c_A \right), & (\sigma > x),
\end{cases}
\label{eq:dydt_piecewise}
\end{align}
where \(b = b_H - b_L\) denotes the benefit differential between the high- and low-quality items.
This piecewise structure arises because the payoff differences are represented differently across the two regimes. For $x \ge \sigma$, active consumers are fully saturated with high-quality items, eliminating any dependence on low-quality allocations. For $x < \sigma$, partial substitution toward low-quality items introduces additional linear terms. 

Equilibrium points are obtained by solving $\dot{x}=\dot{y}=0$.  In this model, the dynamics yield exactly three stable equilibria: a collapse equilibrium at $(0,0)$, a boundary equilibrium at $y=1$ with $0 < x^* < 1$, and a single interior coexistence equilibrium. These equilibria represent potential long-run steady states of the evolutionary dynamics. Local stability was evaluated using the eigenvalues of the Jacobian matrix computed at each equilibrium. The explicit equilibrium coordinates and stability conditions are provided in the SI Appendix, and their qualitative implications are discussed in the Results section.

\section*{Acknowledgments}
This work was supported by the Japan Science and Technology Agency (JST) ERATO Grant Number JPMJER2502.

\bibliographystyle{unsrt}
\bibliography{reference}

\clearpage

\appendix
\section*{Supplymental Materials}
\section{Brief Description of the Proposed Model}

We consider a simple evolutionary model of an attention economy composed of two types of agents: content providers and consumers. 
Each provider chooses between two strategies: producing high-quality content (strategy $H$) or low-quality content (strategy $L$). 
Producing high-quality content is costly, so providers adopting strategy $H$ incur a cost $c_H>0$, whereas those adopting strategy $L$ bear no additional cost. Strategy $H$ represents effort-intensive, high-quality content production.

Consumers also choose between two strategies: actively searching for and consuming high-quality content (active strategy $A$), or randomly consuming available content (passive strategy $P$). 
The active strategy entails a search cost $c_A>0$, reflecting the time and cognitive effort required to identify high-quality content. 
Consumers receive higher benefits from high-quality content, earning $b_H>0$ for high-quality and $b_L>0$ for low-quality content, with $b_H>b_L$. 
Our main question is which strategy profiles become evolutionarily stable under the coupled dynamics of providers and consumers.

Attention is treated as a limited cognitive resource.  
Each consumer can allocate attention to only a fraction $\sigma$ of all available content items, where $0<\sigma<1$, reflecting realistic constraints on time and cognitive capacity in online environments.  
This limitation shapes competition among providers for visibility and influences how consumers evaluate and select content.

We assume a population of mproviders and $n$ consumers.  
At each time step, the following sequence occurs:
\begin{enumerate}
  \item Each provider produces one content item according to their strategy: high-quality under $H$ and low-quality under $L$.
  \item Each consumer selects $m\sigma$ content items from those produced in the current step. Active consumers ($A$) preferentially choose high-quality items, while passive consumers ($P$) select items at random.
  \item Each provider receives a reward $r>0$ for each view of their content.
  \item Each consumer obtains benefit $b_H$ or $b_L$ from the content they consume, depending on its quality.
\end{enumerate}
These steps are iterated over time, and the population shares of the four strategies evolve according to replicator dynamics.

\section{Formulation of the Replicator Dynamics}

Let $x\in[0,1]$ be the proportion of providers adopting the high–effort strategy $H$ (so $1-x$ adopts $L$), and let $y\in[0,1]$ be the proportion of consumers adopting the active strategy $A$ (so $1-y$ adopts $P$).  
We denote by $\pi_H^{\mathrm{prov}}$, $\pi_L^{\mathrm{prov}}$, $\pi_A^{\mathrm{cons}}$, and $\pi_P^{\mathrm{cons}}$ the expected payoffs for the four strategies.

The average payoffs for each population are
\begin{equation}
\bar{\pi}_{\mathrm{prov}}
= x\,\pi_H^{\mathrm{prov}} + (1-x)\,\pi_L^{\mathrm{prov}}, 
\qquad
\bar{\pi}_{\mathrm{cons}}
= y\,\pi_A^{\mathrm{cons}} + (1-y)\,\pi_P^{\mathrm{cons}} .
\end{equation}
The two–population evolutionary dynamics follow the standard replicator equations:
\begin{align}
\dot{x} &= x\,\bigl(\pi_H^{\mathrm{prov}} - \bar{\pi}_{\mathrm{prov}}\bigr)
      = x(1-x)\bigl(\pi_H^{\mathrm{prov}} - \pi_L^{\mathrm{prov}}\bigr), \label{eq:RDE_general1} \\ 
\dot{y} &= y\,\bigl(\pi_A^{\mathrm{cons}} - \bar{\pi}_{\mathrm{cons}}\bigr)
      = y(1-y)\bigl(\pi_A^{\mathrm{cons}} - \pi_P^{\mathrm{cons}}\bigr).
\label{eq:RDE_general2}
\end{align}
The expected payoffs can be expressed in closed form, but their expressions depend on whether the fraction of high-effort providers $x$ is larger or smaller than the attention capacity $\sigma$.  

First, we present the payoff expressions and then summarize the resulting piecewise replicator system. 
Each payoff is derived directly from the expected number of views received by each provider or the expected number of items consumed by each viewer, multiplied by the corresponding reward or benefit. Providers receive a reward $r$ per view and incur a cost only when producing high-quality content. Consumers have $m\sigma$ attention slots, and whether these slots can be fully filled with high-quality items determines the structure of the payoffs.

A key observation is that the system naturally divides at the threshold $\sigma = x$. 
This threshold arises because each viewer has $m\sigma$ attention slots, while the total supply of high-quality items is $mx$; thus, the condition $m\sigma \le mx$ is equivalent to $\sigma \le x$.
When the available high-quality content (proportional to $x$) is sufficient to fill the attention capacity of all active consumers, they allocate their attention entirely to high-quality items. 
Conversely, when high-quality content is scarce ($\sigma > x$), active consumers must supplement their attention with low-quality content, leading to different mixing proportions across content types. 
This shift in attention allocation produces two distinct regimes, each with its own payoff expressions.

In both regimes, payoff terms reflect the distribution of attention across content types. 
Passive consumers sample content uniformly at random; therefore, the expected fraction of high-quality items they consume is simply $x$. 
Active consumers, in contrast, exhaust the available supply of high-quality items before allocating attention elsewhere.

In the regime $\sigma \le x$, their attention can be fully concentrated on high-quality content. In the regime $\sigma > x$, only a fraction $x/\sigma$ of attention can be allocated to high-quality items, and the remainder must be filled with low-quality items. 
For providers, the term $(1-y)$ captures the contribution of passive consumers, who allocate their $\sigma m$ attention slots uniformly across all $m$ items. 
The term $y/x$ arises from active consumers: there are $ny$ active viewers, each selecting $\sigma m$ items, so the total number of active views on high-quality content is $ny\sigma m$. Dividing this by the $mx$ high–quality items yields an expected $ny\sigma/x$ views per high-quality item; after factoring out $rn\sigma$ in the payoff expression, this appears as the factor $y/x$. 
In regime $\sigma \le x$, the resulting payoffs are
\begin{align}
\pi_H^{\mathrm{prov}}(x,y) &= -c_H + r n \sigma\left((1-y) + \frac{y}{x}\right),\\
\pi_L^{\mathrm{prov}}(x,y) &= r n \sigma(1-y),\\
\pi_A^{\mathrm{cons}}(x,y) &= -c_A + m\sigma b_H,\\
\pi_P^{\mathrm{cons}}(x,y) &= \sigma m\left(x b_H + (1-x)b_L\right).
\end{align}

In the regime $\sigma > x$, active consumers first exhaust all available high-quality items. Consequently, every high-quality item is viewed by all active consumers, and the total attention allocated to high-quality content is $n y m x$. The remaining attention, $n y m(\sigma - x)$, must be allocated to low-quality items. Since there are $m(1-x)$ low-quality items, this residual attention is distributed uniformly across them, yielding an average of $(\sigma - x)/(1-x)$ views per low-quality item for each active consumer. 
Therefore, in regime $\sigma > x$, the payoffs become
\begin{align}
\pi_H^{\mathrm{prov}}(x,y) &= -c_H + r n\left(\sigma(1-y) + y\right),\\
\pi_L^{\mathrm{prov}}(x,y) &= r n\left[\sigma(1-y) + y\frac{\sigma-x}{1-x}\right],\\
\pi_A^{\mathrm{cons}}(x,y) &= -c_A + m\left(x b_H + (\sigma-x)b_L\right),\\
\pi_P^{\mathrm{cons}}(x,y) &= \sigma m\left(x b_H + (1-x)b_L\right).
\end{align}

For the subsequent stability analysis, we define the effective parameter $b := b_H - b_L > 0$ and incorporates $b_L$ into the baseline payoff.
With this substitution, the replicator dynamics can be expressed in the simplified piecewise system:
\begin{align}
\frac{dx}{dt} &=
\begin{cases}
(1 - x)\left( r n \sigma y - c_H x \right), & (\sigma \leq x), \\[0.5em]
x \left( r n y(1 - \sigma) + c_H(x - 1) \right), & (\sigma > x),
\end{cases} 
\label{eq:dxdt_piecewise}
\\[0.5em]
\frac{dy}{dt} &=
\begin{cases}
y(1 - y)\left( b m \sigma (1 - x) - c_A \right), & (\sigma \leq x), \\[0.5em]
y(1 - y)\left( b m x (1 - \sigma) - c_A \right), & (\sigma > x),
\end{cases}
\label{eq:dydt_piecewise}
\end{align}
where $r, n, m, b, c_H, c_A > 0$ are constant parameters and $0<\sigma<1$ is the attention ratio.

\section{Derivation and Stability of Equilibria}

The equilibria are determined by the simultaneous solutions of $\dot{x}=0$ and $\dot{y}=0$. Once these fixed points are identified, their stability can be assessed by examining the evolution of small perturbations under the dynamics. Since the replicator system is two-dimensional and nonlinear, local stability is evaluated by linearizing the system around each equilibrium.

Let the system be expressed as $\dot{x}=f(x,y)$ and $\dot{y}=g(x,y)$. 
The Jacobian matrix
\[
J(x,y)=
\begin{bmatrix}
\frac{\partial f}{\partial x} & \frac{\partial f}{\partial y} \\
\frac{\partial g}{\partial x} & \frac{\partial g}{\partial y}
\end{bmatrix}
\]
captures the local behavior of trajectories near $(x,y)$. 
Evaluating $J(x,y)$ at equilibrium $(x^*,y^*)$ provides a linear approximation of the dynamics.  
The equilibrium is
\begin{itemize}
    \item locally asymptotically stable if both eigenvalues of $J(x^*,y^*)$ have negative real parts,
    \item unstable if at least one eigenvalue has a positive real part.
\end{itemize}

Because the payoff functions that determine $f$ and $g$ differ depending on whether $\sigma \le x$ or $\sigma > x$, 
Partial derivatives must be computed separately for these two regions.  
Accordingly, the correct branch of the Jacobian must be used to evaluate the stability of each equilibrium.  
If a candidate equilibrium lies at the switching boundary $x=\sigma$, the Jacobian should be evaluated using the appropriate one-sided limit.

For reference, we summarize the Jacobian matrix in these two regions. 
From Eqs. \eqref{eq:dxdt_piecewise}–\eqref{eq:dydt_piecewise}, we obtain the Jacobian matrix
\begin{equation}
J(x, y) =
\begin{bmatrix}
- rn\sigma y + c_H(2x - 1) & (1 - x) rn\sigma \\
- y(1 - y) bm\sigma & (1 - 2y)(bm\sigma(1 - x) - c_A)
\end{bmatrix}
\quad (\sigma\le x),
\label{J_sigma_x}
\end{equation}
and
\begin{equation}
J(x, y) =
\begin{bmatrix}
rny(1 - \sigma) + c_H(2x - 1) & x\, rn(1 - \sigma) \\
y(1 - y)\, bm(1 - \sigma) & (1 - 2y)(bmx(1 - \sigma) - c_A)
\end{bmatrix}
\quad (\sigma>x).
\end{equation}
\label{J_x_sigma}

\subsection{Corner Equilibria}

The state space $[0,1]^2$ has the vertexes $(x, y) \in \{(0,0), (0,1), (1,0), (1,1)\}$.
Each represents a homogeneous population of providers and consumers.
Because the replicator equations-- Eqs. \eqref{eq:RDE_general1}-\eqref{eq:RDE_general2}-- have the factors of $x(1-x)$ and $y(1-y)$, all four corners automatically satisfy $\dot{x}=\dot{y}=0$.

Stability is determined by evaluating the Jacobian matrix $J(x,y)$
introduced above. Since the Jacobian is piecewise, an appropriate
branch must be selected based on whether $\sigma \le x$ or $\sigma > x$.

\paragraph{(i) $(x,y)=(0,0)$.}
Here, $x<\sigma$. Thus, we use the Jacobian for $\sigma > x$.
Substituting $(0,0)$ yields
\begin{equation}
J(0,0)=
\begin{bmatrix}
- c_H & 0\\
0 & - c_A
\end{bmatrix},
\end{equation}
whose eigenvalues $(-c_H,\,-c_A)$ are both negative.
Therefore, $(0,0)$ is always asymptotically stable.

\paragraph{(ii) $(x,y)=(0,1)$.}
Again, $x<\sigma$; hence, the branch $\sigma > x$ applies:
\begin{equation}
J(0,1)=
\begin{bmatrix}
rn(1-\sigma)-c_H & 0\\
0 & c_A
\end{bmatrix}.
\end{equation}
Because $c_A>0$, this point is unstable.

\paragraph{(iii) $(x,y)=(1,0)$.}
Here, $x=1 \ge \sigma$. Thus, we use the branch $\sigma \le x$:
\begin{equation}
J(1,0)=
\begin{bmatrix}
c_H & 0\\
0 & -c_A
\end{bmatrix}.
\end{equation}
One eigenvalue is positive and one is negative; thus, $(1,0)$ is unstable.

\paragraph{(iv) $(x,y)=(1,1)$.}
Again, $\sigma \le x$:
\begin{equation}
J(1,1)=
\begin{bmatrix}
c_H - rn\sigma & 0\\
0 & c_A
\end{bmatrix}.
\end{equation}
Because $c_A>0$, the point is unstable, regardless of the sign of $c_H - rn\sigma$.

Hence, among the corner equilibria, only $(0,0)$ is asymptotically stable and the remaining three corners are unstable.

\subsection{Boundary Behavior on $x=0,1$ and $y=0$}
Next, we examined the equilibria that lie on the edges of the state space.  
Each boundary of $[0,1]^2$ is invariant under the dynamics, so the flow along an edge reduces to a one-dimensional equation.  
Before analyzing the nontrivial case $y=1$, we first summarize the behavior along the other boundaries $x=0$, $x=1$, and $y=0$.

\paragraph{Boundary $x=0$.}

For $x=0$ we have $x<\sigma$; therefore, the case $\sigma > x$ in Eqs.
\eqref{eq:dxdt_piecewise}–\eqref{eq:dydt_piecewise} applies:
\begin{equation}
\dot{x} = 0,\qquad
\dot{y} = y(1-y)\bigl(bm\sigma - c_A\bigr).
\end{equation}
Thus, along $x=0$ the system reduces to one-dimensional dynamics in $y$, and the only equilibria on this edge are the corners $(0,0)$ and $(0,1)$.

\paragraph{Boundary $x=1$.}

For $x=1$, $\sigma \le x$. Hence, 
\begin{equation}
\dot{x} = 0,\qquad
\dot{y} = y(1-y)\bigl(bm\sigma(1-1) - c_A\bigr)
= -c_A\,y(1-y).
\end{equation}
Again, the dynamics are one-dimensional in $y$, and the only equilibria on $x=1$ are the corners $(1,0)$ and $(1,1)$.

\paragraph{Boundary $y=0$.}

For $y=0$, we have $\dot{y}=0$ and
\begin{equation}
\dot{x} =
\begin{cases}
(1-x)(0 - c_H x) = -c_H x(1-x), & (\sigma \le x), \\[0.3em]
x\bigl(c_H(x-1)\bigr), & (\sigma > x),
\end{cases}
\end{equation}
such that $\dot{x}<0$ holds for all $0<x<1$. Along $y=0$ the flow runs from $(1,0)$ to $(0,0)$, and the only equilibria on this edge are the corners $(0,0)$ and $(1,0)$.

In summary, there are no nontrivial equilibria along the boundaries $x=0$, $x=1$, or $y=0$, and all boundary equilibria other than the four corners must lie on $y=1$. Therefore, we next focus on the line $y=1$ to analyze the existence and stability of interior points on this boundary.

\subsection{Boundary Equilibria on $y=1$}
Finally, we examine the boundary $y=1$, which is the only edge that may contain nontrivial equilibria.  
By setting $y=1$ in Eq. \eqref{eq:dxdt_piecewise}, for $\sigma \le x$:
\begin{equation}
      \dot{x} = (1 - x)(rn\sigma - c_H x),
\end{equation}
  such that the nontrivial equilibria satisfy
\begin{equation}
  x^* = \frac{rn\sigma}{c_H}.
\end{equation}
For $\sigma > x$:
\begin{equation}
  \dot{x} = x\big(rn(1 - \sigma) + c_H(x - 1)\big),
\end{equation}
such that the nontrivial equilibria satisfy
\begin{equation}
  x^* = 1 - \frac{rn(1 - \sigma)}{c_H}.
\end{equation}

Thus, $y=1$ may host two candidate boundary equilibria
\begin{equation}
(x^*, y^*) = \left(\frac{rn\sigma}{c_H},\, 1\right), 
\qquad
(x^*, y^*) = \left(1 - \frac{rn(1 - \sigma)}{c_H},\, 1\right),
\end{equation}
provided they fall within $(0,1)$ and satisfy the corresponding regional conditions for $x$.

\subsubsection*{Stability of $(x^*,y^*)=(\frac{rn\sigma}{c_H},1)$ in  $\sigma\le x$}

For $\sigma\le x$, we use the following Jacobian for $\sigma\le x$ evaluated at $(x^*,1)$:
\begin{equation}
J(x^*, 1) =
\begin{bmatrix}
- rn\sigma + c_H(2x^* - 1) & (1 - x^*) rn\sigma \\
0 & - (bm\sigma(1 - x^*) - c_A)
\end{bmatrix}.
\end{equation}
This matrix is upper-triangular; therefore, the eigenvalues are diagonal. Substituting $x^* = rn\sigma/c_H$ yields
\begin{align*}
\lambda_1 &= - rn\sigma + c_H\left(2\frac{rn\sigma}{c_H} - 1\right)
= rn\sigma - c_H, \\
\lambda_2 &= -\left(bm\sigma\left(1 - \frac{rn\sigma}{c_H}\right) - c_A\right)
= c_A - bm\sigma\left(1 - \frac{rn\sigma}{c_H}\right).
\end{align*}
Thus, $(x^*,1)$ is asymptotically stable if and only if
\begin{align}
c_H &> rn\sigma, \label{eq:y1_stable_1}\\
c_A &< bm\sigma\left(1 - \frac{rn\sigma}{c_H}\right). \label{eq:y1_stable_2}
\end{align}

Note that the existence of this point as a nontrivial equilibrium with $\sigma < x^* < 1$ requires
\begin{equation}
rn\sigma < c_H < rn.
\end{equation}
Combined with \eqref{eq:y1_stable_2}, a convenient form of the stability condition is as follows:
\begin{equation}
rn\sigma < c_H < rn, 
\qquad 
\frac{c_A}{bm\sigma} + \frac{rn\sigma}{c_H} < 1.
\end{equation}

\subsubsection*{Stability of $(x^*,y^*)=(1-\frac{rn(1-\sigma)}{c_H},1)$ in $\sigma>x$}

For $\sigma>x$, we used the Jacobian with $\sigma>x$:
\begin{equation}
J(x^*, 1) =
\begin{bmatrix}
rn(1 - \sigma) + c_H(2x^* - 1) & x^* rn(1 - \sigma) \\
0 & - (bmx^*(1 - \sigma) - c_A)
\end{bmatrix}.
\end{equation}
Substituting $x^* = 1 - \frac{rn(1 - \sigma)}{c_H}$ yields
\begin{align*}
\lambda_1 &= c_H - rn(1 - \sigma), \\
\lambda_2 &= c_A - bm(1 - \sigma)\left(1 - \frac{rn(1 - \sigma)}{c_H}\right).
\end{align*}
For $(x^*,1)$ to be in the region of $\sigma>x$, we require $0<x^*<\sigma$, which is equivalent to
\begin{equation}
0<1 - \frac{rn(1-\sigma)}{c_H} < \sigma
\quad\Rightarrow\quad
0<c_H-rn(1-\sigma)<c_H\sigma.
\end{equation}
However, this implies that $\lambda_1 = c_H - rn(1-\sigma) > 0$. Thus, $(x^*,1)$ cannot be stable in its defined region. Therefore, the equilibrium $(1 - \frac{rn(1-\sigma)}{c_H},1)$ is always unstable.

Thus, the only possible stable equilibrium for $y=1$ is
\begin{equation}
(x^*, y^*) = \left(\frac{rn\sigma}{c_H},\, 1\right),
\end{equation}
and it is stable for
\begin{equation}
rn\sigma < c_H < rn, 
\qquad 
\frac{c_A}{bm\sigma} + \frac{rn\sigma}{c_H} < 1.
\end{equation}

\subsection{Interior Equilibria}

We now consider the equilibria for $0<x<1$ and $0<y<1$.

\subsubsection*{Case $\sigma \le x$}

From Eqs. \eqref{eq:dxdt_piecewise}–\eqref{eq:dydt_piecewise}, interior equilibria in $\sigma\le x$ satisfy
\begin{align*}
(1 - x)(rn\sigma y - c_H x) &= 0 
\quad\Rightarrow\quad 
y = \frac{c_H x}{rn\sigma}, \\[0.25em]
y(1 - y)\big(bm\sigma(1 - x) - c_A\big) &= 0 
\quad\Rightarrow\quad 
x = 1 - \frac{c_A}{bm\sigma}.
\end{align*}
Thus, the interior equilibrium is
\begin{equation}
(x^*, y^*) = 
\left(
1 - \frac{c_A}{bm\sigma},\ 
\frac{c_H}{rn\sigma}\left(1 - \frac{c_A}{bm\sigma}\right)
\right).
\end{equation}
For $0<\sigma<1$ and suitable choices for $c_A$ and $c_H$, 
the resulting point $(x^*,y^*)$ satisfies $\sigma < x^* < 1$ and $0<y^*<1$.
The precise existence conditions are summarized below:
\begin{equation}
    bm\sigma(1-\sigma) > c_A, \qquad \frac{c_A}{bm\sigma} + \frac{rn\sigma}{c_H} > 1.
\end{equation}

Evaluating the Jacobian for $\sigma\le x$ at $(x^*,y^*)$ and simplifying it, we obtain
\begin{equation}
J =
\begin{bmatrix}
- c_H(1-x^*) & rn\sigma (1-x^*) \\
- bm\sigma y^*(1-y^*) & 0
\end{bmatrix}.
\end{equation}
The trace and determinant of the matrix are as follows:
\begin{equation}
\operatorname{tr}(J) = -c_H(1-x^*) < 0, 
\qquad 
\det(J) = - J_{12}J_{21} > 0
\end{equation}
because $J_{12}>0$ and $J_{21}<0$ for $0<y^*<1$.
Therefore, both eigenvalues have negative real parts, and we conclude that the interior equilibrium $(x^*,y^*)$ in region $\sigma\le x$ is always asymptotically stable if it exists. The existence conditions $x^*\in(\sigma,1)$ and $y^*\in(0,1)$ can be summarized as follows:
\begin{equation}
bm\sigma(1-\sigma) > c_A,
\qquad
\frac{c_A}{bm\sigma} + \frac{rn\sigma}{c_H} > 1.
\end{equation}

\subsubsection*{Case $\sigma > x$}

In this region, the interior equilibrium satisfies
\begin{align*}
x\big(rny(1 - \sigma) + c_H(x - 1)\big) &= 0
\quad\Rightarrow\quad
y = \frac{c_H(1 - x)}{rn(1 - \sigma)}, \\[0.25em]
y(1 - y)\big(bmx(1 - \sigma) - c_A\big) &= 0
\quad\Rightarrow\quad
x = \frac{c_A}{bm(1 - \sigma)}.
\end{align*}
Thus, the interior equilibrium is
\begin{equation}
(x^*, y^*) = 
\left(
\frac{c_A}{bm(1 - \sigma)},\ 
\frac{c_H}{rn(1 - \sigma)}\left(1 - \frac{c_A}{bm(1 - \sigma)}\right)
\right),
\end{equation}
with $0<x^*,y^*<1$ and $\sigma > x^*$ assumed.

In this case, the Jacobian is simplified to
\begin{equation}
J =
\begin{bmatrix}
c_H x^* & J_{12} \\
J_{21} & 0
\end{bmatrix},
\end{equation}
where
\begin{align*}
J_{12} &= x^* rn(1 - \sigma) = \frac{rn c_A}{bm} > 0, \\
J_{21} &= y^*(1 - y^*) bm(1 - \sigma) > 0.
\end{align*}
Thus
\begin{equation}
\operatorname{tr}(J) = c_H x^* > 0,
\qquad
\det(J) = -J_{12}J_{21} < 0.
\end{equation}
Therefore, the eigenvalues have opposite signs, and we conclude that the interior equilibrium in the region $\sigma > x$ is always unstable.

\subsection{Summary of Stable Equilibria}

In summary, the system admits the following stable equilibria:

\begin{itemize}
  \item[(1)] The collapse equilibrium
\begin{equation}
  (x, y) = (0, 0),
\end{equation}
  which is always asymptotically stable 
  ($\lambda_1 = -c_H < 0$, $\lambda_2 = -c_A < 0$).

  \item[(2)] A boundary equilibrium on $y=1$:
\begin{equation}
  (x, y) = \left(\frac{rn\sigma}{c_H},\, 1\right)
  \end{equation}
  with $x\in(\sigma,1)$, which is stable if and only if
  \begin{equation}
  rn\sigma < c_H < rn, 
  \qquad 
  \frac{c_A}{bm\sigma} + \frac{rn\sigma}{c_H} < 1.
  \end{equation}

  \item[(3)] An interior equilibrium in the region $\sigma\le x$:
\begin{equation}
  (x, y) = \left(
  1 - \frac{c_A}{bm\sigma},\ 
  \frac{c_H}{rn\sigma}\left( 1 - \frac{c_A}{bm\sigma} \right)
  \right),
\end{equation}
  with $\sigma < x < 1$ and $0<y<1$, which is always asymptotically stable whenever it exists. The existence conditions can be written as
  \begin{equation}
  bm\sigma(1-\sigma) > c_A,
  \qquad
  \frac{c_A}{bm\sigma} + \frac{rn\sigma}{c_H} > 1.
  \end{equation}
\end{itemize}

\section{Boundary Between Collapse and Non–Collapse Regimes}

Finally, we clarify the conditions for a purely collapsing system, in which the only stable equilibrium is $(x,y)=(0,0)$. In the classification above, the collapse equilibrium $(0,0)$ is always locally stable, whereas additional stable equilibria may exist in certain regions of parameter space. In particular, nontrivial stable equilibria (on $y=1$ or in the interior) require
\begin{equation}
bm\sigma(1-\sigma) > c_A
\end{equation}
along with the inequalities involving $c_H$.

For a stable boundary equilibrium at $y=1$, we require
\begin{equation}
  rn\sigma < c_H < rn, 
  \qquad 
  \frac{c_A}{bm\sigma} + \frac{rn\sigma}{c_H} < 1.
\end{equation}
  From $rn\sigma < c_H < rn$ we obtain:
\begin{equation}
  1 - \frac{rn\sigma}{c_H} > 1 - \frac{rn\sigma}{rn} = 1 - \sigma,
\end{equation}
  and hence
\begin{equation}
  c_A < bm\sigma\left(1 - \frac{rn\sigma}{c_H}\right) < bm\sigma(1-\sigma).
\end{equation}

  Thus, the condition $bm\sigma(1-\sigma) > c_A$ is automatically satisfied when the boundary equilibrium is stable.

For a stable interior equilibrium in $\sigma\le x$, we explicitly require
\begin{equation}
  bm\sigma(1-\sigma) > c_A,
  \qquad
  \frac{c_A}{bm\sigma} + \frac{rn\sigma}{c_H} > 1.
\end{equation}

The parameter region in which the system admits nontrivial stable equilibria (either on $y=1$ or in the interior) is contained within the domain $bm\sigma(1-\sigma) > c_A$.
Conversely, the complementary inequality $bm\sigma(1-\sigma) < c_A$
implies that no additional stable equilibria exist, and the only stable equilibrium is the collapse equilibrium $(x,y)=(0,0)$. Thus, the equation
\begin{equation}
    bm\sigma(1 -\sigma) = c_A
\end{equation}

defines the boundary that separates the collapse regime (where only $(0,0)$ is stable) from the coexistence regime, where high-quality strategies can persist.

\end{document}